\documentclass[runningheads]{llncs}
\usepackage{graphicx}
\usepackage{xcolor}
\usepackage{float}
\begin{document}
\title{In-silico analysis of the influence of pulmonary vein configuration on left atrial haemodynamics and thrombus formation in a large cohort}
%
%\titlerunning{Abbreviated paper title}
% If the paper title is too long for the running head, you can set
% an abbreviated paper title here
%
\author{Jordi Mill\inst{1} \and Josquin Harrison\inst{2} \and Benoit Legghe\inst{3} \and Andy L. Olivares\inst{1}  \and Xabier Morales\inst{1} \and Jerome Noailly\inst{1} \and Xavier Iriart\inst{3} \and Hubert Cochet\inst{3} \and Maxime Sermesant\inst{2} \and Oscar Camara\inst{1} } 
\authorrunning{J. MIll et al.}
% First names are abbreviated in the running head.
% If there are more than two authors, 'et al.' is used.
%
\institute{Physense, BCN Medtech, Department of Information and Communication Technologies, Universitat Pompeu Fabra, Barcelona, Spain\\ \email{jordi.mill@upf.edu} \and
Inria, Université Côte d'Azur, Epione team, Sophia Antipolis, France\and H\^{o}pital de Haut-L\'{e}v\^{e}que, Bordeaux, France\\
}
\maketitle              % typeset the header of the contribution
\begin{abstract}
Atrial fibrillation (AF) is considered the most common human arrhythmia. Around 99\% of thrombi in non-valvular AF are formed in the left atrial appendage (LAA). Studies suggest that abnormal LAA haemodynamics and the subsequently stagnated flow are the factors triggering clot formation.  However, the relation between LAA morphology, the blood pattern and the triggering is not fully understood. Moreover, the impact of structures such as the pulmonary veins (PVs) on LA haemodynamics has not been thoroughly studied due to the difficulties of acquiring appropriate data. On the other hand, in-silico studies and flow simulations allow a thorough analysis of haemodynamics, analysing the 4D nature of blood flow patterns under different boundary conditions. However, the reduced number of cases reported on the literature of these studies has been a limitation. The main goal of this work was to study the influence of PVs on left atrium (LA) and LAA haemodynamics. Computational fluid dynamics simulations were run on 52 patients, the largest cohort so far in the literature, where different parameters were individually studied: pulmonary veins orientation and configuration; LAA and LA volumes and its ratio; and flow velocities. Our computational analysis showed how the right pulmonary vein height and angulation have a great influence on LA haemodynamics. Additionally, we found that LAA with great bending with its tip pointing towards the mitral valve could contribute to favour flow stagnation. 

\end{abstract}
\section{Introduction}
Atrial fibrillation (AF) is considered the most common of human arrhythmias. Approximately 2\% of people younger than age 65 have AF, rising to about 9\% of people aged 65 years or more \cite{Rahman2014}. AF is currently seen as a marker of an increased risk of stroke since it favours thrombus formation inside the left atrium (LA). Around 99\% of thrombi in non-valvular AF are formed in the left atrial appendage (LAA) \cite{Cresti2019}. LAA shapes are complex and have a high degree of anatomical variability among the population. Thus, researchers have sought to classify LAA morphologies and relate them to the risk of thrombus formation \cite{DiBiase2012}. However, no classification has achieved a consensus due to their subjective interpretation and contradictions in their relationship with thrombus formation. Blood flow hemodynamics is another relevant factor for thrombogenesis, following Virchow’s triad principles \cite{Watson2009}; low velocities and stagnated flow have been associated with the triggering of the inflammatory process and, therefore, the risk of thrombus generation \cite{Naser2011}. 

The pulmonary veins’ (PVs) configuration and orientation play a key role in radiofrequency ablation therapy (e.g., PVs isolation) since they are a preferential origin of ectopic foci in atrial fibrillation. On the other hand, there are not robust studies on the relation between the PVs configuration and the hemodynamics in the LA, including the LAA and, as a consequence, the thrombus formation process \cite{CRONIN2007}. Some large-scale studies have classified PVs configurations into different anatomical categories \cite{marom2004} but they have never been related with haemodynamics and thrombus formation. Moreover, there is high anatomical variability among the population, where most of humans have 4 PVs but there are reported cases of 3, 5, 6 or even 7 PVs present. Even more, the orientation of how the PVs are inserted into the LA can differ substantially from patient to patient.

In daily clinical practice, LA haemodynamics is mainly studied using echocardiographic images, usually simplified to a single blood flow velocity value at one point in space and time (e.g. center of LAA ostium at end-diastole)[8]. Advanced imaging techniques such as 4D flow Magnetic Resonance Images (MRI), allowing a more complete blood flow analysis, are emerging but they still provide limited information in the left atria \cite{Beigel2014}. Therefore, the question if the PVs configuration changes LA hemodynamics remains open. Recently \cite{Markl2015}, researchers have investigated the dependence of blood flow entering to the LAA and related thrombogenic indices with different PVs configurations, but only in synthetic cases. At this juncture, patient-specific  models based on computational fluid dynamics can provide a better haemodynamic characterization of the LA and LAA, deriving in-silico indices of the blood flow at each point of the geometry over time. In the last decade there have been several attempts to develop simulation frameworks for the blood flow analysis of the human LA and LAA  \cite{Guadalupe2018}, but only applied to a very limited number of patient-specific cases ($<$ 10) and independently of morphological parameters \cite{Fumagalli2020,Otani2016}. In this study, we built patient-specific computational models of 52 patients with atrial fibrillation and used computational fluid dynamics (CFD) simulations to study their hemodynamics in relation with PVs configurations, constituting the largest cohort of LA-based fluid simulations in the literature. 

\section{Methods}
\subsection{Clinical data}
The clinical data used in this work were provided by Hospital Haut-Lévêque (Bordeaux, France), including AF patients that underwent a left atrial occlusion (LAAO) intervention and with available pre-procedural high-quality Computed Tomography (CT) scans. 52 patients were selected for this study. During the data processing pipeline, clinical decisions and patient outcomes were hidden to the researchers setting up the in-silico simulations. Cardiac CT studies were performed on a 64-slice dual source CT system (Siemens Definition, Siemens Medical Systems, Forchheim, Germany). Images were acquired using a biphasic injection protocol: 1 mL/kg of Iomeprol 350 mg/mL (Bracco, Milan, Italy) at the rate of 5 mL/s, followed by a 1 mL/kg flush of saline at the same rate. 31 CT images were acquired in systole (15 with stroke history) and 21 in diastole (10 with stroke history). The study was approved by the Institutional Ethics Committee, and all patients provided informed consent.

Regarding PVs configuration the subjects were grouped as follows: 2 cases with 3 PVs; 23 with 4 PVs; 19 with 5 PVs; 6 with 6 PVs; and 2 with 7 PVs. For LAA volume, the distribution of the patients stands as follows: 4 cases between 1-5 mL; 14 between 10-15 mL; 18 between 10-15 mL; 12 between 15-20 mL; and 4 cases with more than 20 mL. Regarding thrombus formation, 27 cases were in the control group while 25 had a history of stroke or a thrombus was found in the LAA. 

\subsection{In-silico computational model}
The LA geometries were segmented from the CT images by a different member who did not participate in the subsequently modelling process in order to maintain the study blind to the modellers. The final volumetric meshes were between 8 - 9 $\times 10^5$ elements, depending on the volume of the LA after performing a mesh convergence study up to 1M elements.

A velocity curve at the mitral valve (MV) was obtained from Doppler ultrasound from a patient of the whole database to impose as outlet boundary condition. The same pressure wave from an AF patient was used as inlet boundary conditions at the pulmonary veins. The movement of the mitral valve ring plane was defined according to literature\cite{Veronesi2008} and was diffused through the whole LA with a dynamic mesh approach based on the spring based method implemented on the CFD solver in Ansys Fluent 19 R32 (ANSYS Inc, USA). The edges between any two nodes were idealised as a network of springs and the Hooke's Law was applied in each node. Physiologically speaking, the method tries to mimic the longitudinal movement passively produced in the LA by the contraction of the left ventricle (LV). The lack of radial movement tries to mimic the lack of active contraction since the patient suffers from AF. 
All boundary conditions were synchronised with the patient’s electrocardiogram. . Post-processing and visualisation of simulation results were performed using ParaView 5.4.13 \footnote{https://www.paraview.org/} . The blood was treated as Newtonian fluid, with a density of 1060 Kg/m3 and a viscosity of 0.0035 Pa/s. Three heart beats were calculated with a 0.01 s time step size. 

\subsection{Haemodynamic descriptors}

To assess the origin of the flow entering the LAA, 50 seeds were placed at the LAA, from which streamlines were computed. A streamline can be thought of as the path a mass-less particle takes flowing through a velocity field (i.e., vector field) at a given instant in time. In order to have a complete understanding of the blood flow patterns under study, the streamlines were computed in different time frames over the heartbeat in order to gather more representative samples. Specifically, the chosen time frames in the cardiac cycle were the following: 1) at the beginning of the ventricular systole; 2) just before MV opens; 3) when the maximum velocity is reached within the E wave (early diastole, after MV opens); 4) middle time point between the E and A wave; and 5) when the maximum velocity of the A wave is reached (end diastole, just before MV closes to start the cycle again). This process was repeated for three beats. The streamlines were also used to localise the position where the main collision between the PVs flows was produced (e.g., from the right and left PVs), that is the moment that the flow coming from each PVs crosses with each other (see Figure 2). Furthermore, blood flow patterns within the LA were studied placing additional 50 streamline seeds in each PVs. 

\begin{figure}[t]
\includegraphics[scale=0.25]{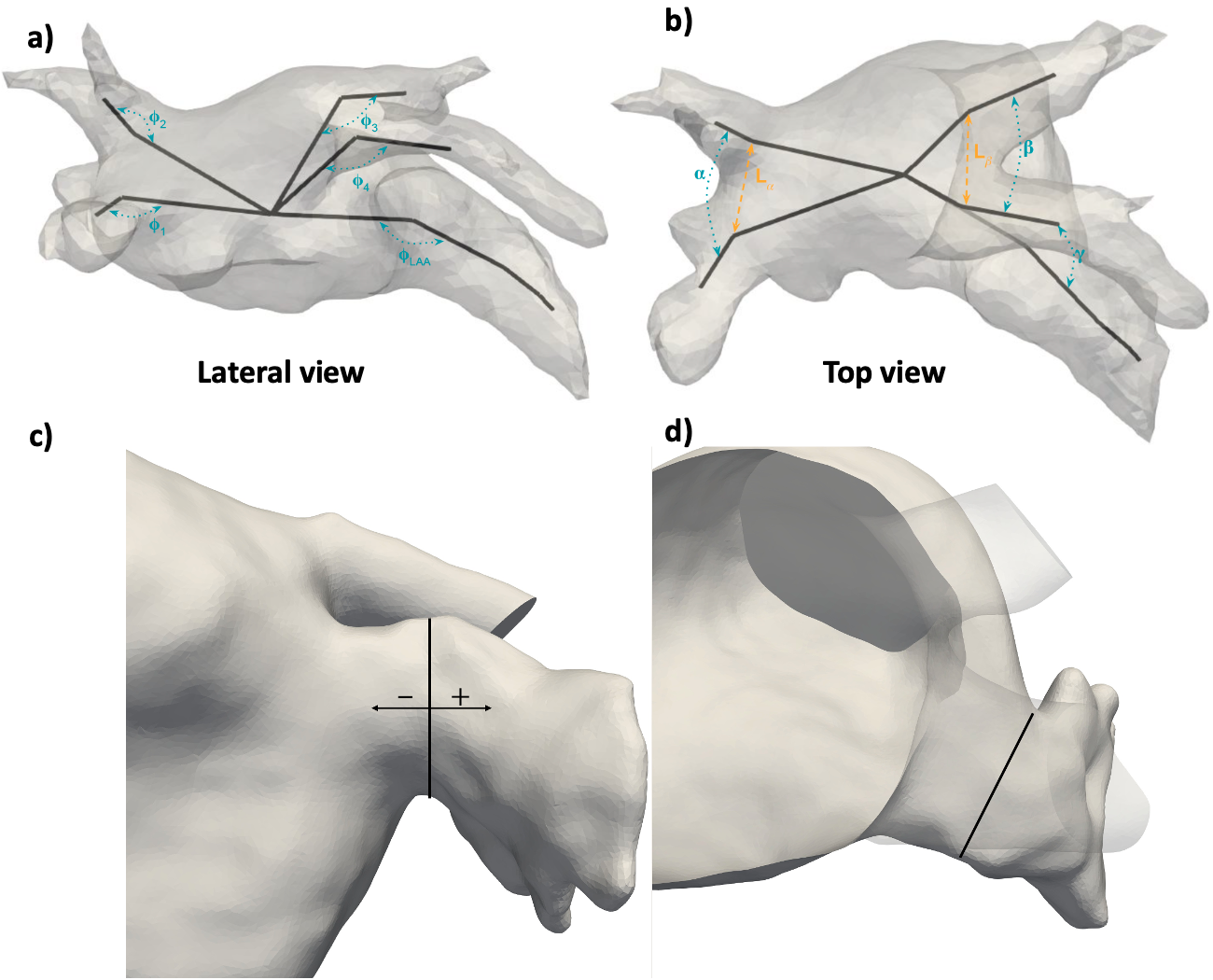}
\centering
\caption{a) Lateral view of the left atrium with the diffrerent $\phi$ angles computed; b)Angles $\alpha$, $\beta$, $\gamma$ and the distances $L_{\alpha}$ and $L_{\beta}$;c) 2D plane (black line) for the flow rate computation, being placed below the first lobe of the left atrial appendage (LAA) bending. Flow was considered positive if entering the LAA and negative, otherwise.} \label{fig1}
\end{figure}

The flow rate was measured from the fluid simulations at the entrance of the LAA, with a robust criterion considering the large anatomical variability. We selected a 2D plane below the first lobe before the LAA bending (see Figure \ref{fig1}), considering flow entering the LAA as positive (and negative with outgoing flow). Two beats were analysed since the first was used to reach convergence and a more stable flow solution. 
 
Subsequently, the obtained flow rate curve was integrated over time to compute the final volume crossing the selected 2D plane. A zero value in the integration would mean that all flow entering the LAA leaves at the end of the beats. On the other hand, a large positive value would indicate that a lot of flow goes inside the LAA without leaving, i.e., potentially signalling flow stagnation. In order to consider the large LAA volume variation in the population, the obtained values were estimated as a percentage of the LAA volume for each case, i.e., obtaining the amount of flow volume staying in the LAA with respect to its volume.

\subsection{Morphological descriptors}

To extract morphological features from the LA, the LAA and each pulmonary vein were labelled;  their barycenter was computed together with the center of intersection of each label, producing a two-point representation of each veins (which we will call a branch). Additionally, for the LAA we added one more point: the LAA was cut at the barycenter with the  normal plane to the LAA centerline; , followed by the computation of the barycenter of the outermost half. Finally, we added the barycenter of the body of LA (the part that was left unlabelled). Overall, we obtained a \textit{skeleton} view of the LA as shown in black in Figure~\ref{fig1}). 
To labelise every mesh, we first applied a diffeomorphic registration (using the deformetrica software\cite{Deformetrica}) from a chosen template shape to the rest of the population. Thanks to this we could transfer the labels so the skeleton representation can be extracted. T
From the points we choose to compute the following morphological features: 
\begin{itemize}
    \item The angle $\alpha$ is the angle between the right inferior pulmonary vein (RIPV), to and the right superior pulmonary vein (RSPV);
    \item Angle $\beta$, equivalent to angle alpha but for the left superior pulmonary vein (LSPV) and left inferior pulmonary vein (LIPV);
    \item $L_{\alpha}$ is the length between RSPV ostium and RIPVs ostium;
    \item $L_{\beta}$ is the length between LSPV ostium and LIPVs ostium;
    \item $\alpha/\beta$ and $L_{\alpha}/L_{\beta}$ being their respective ratios;
    \item Angle $\phi$ is the angle at the PVs intersection between the centre of the LA cavity and the PV centre, characterising the amount of fold between the vein and the rest of the LA;
    \item $\phi_{LAA}$ measures the LAA bending angle described at the middle point in the LAA with respect to its ostium and its tip; 
    \item $\gamma$ measures the angle between the LSPVs and the LAA. To compute it, we perform a rigid transformation of the LSPVs branch to the LAA to have common ostium point and compute the angle at this ostium with respect to the translated point and the middle point of the LAA.   
\end{itemize}

\section{Results}
A thorough qualitative analysis of the streamlines’ patterns, over the three beats and for all cases, showed that the haemodynamics of the LA changed a lot depending on the PVs configuration. 

\begin{figure}[h]
\includegraphics[scale=0.4]{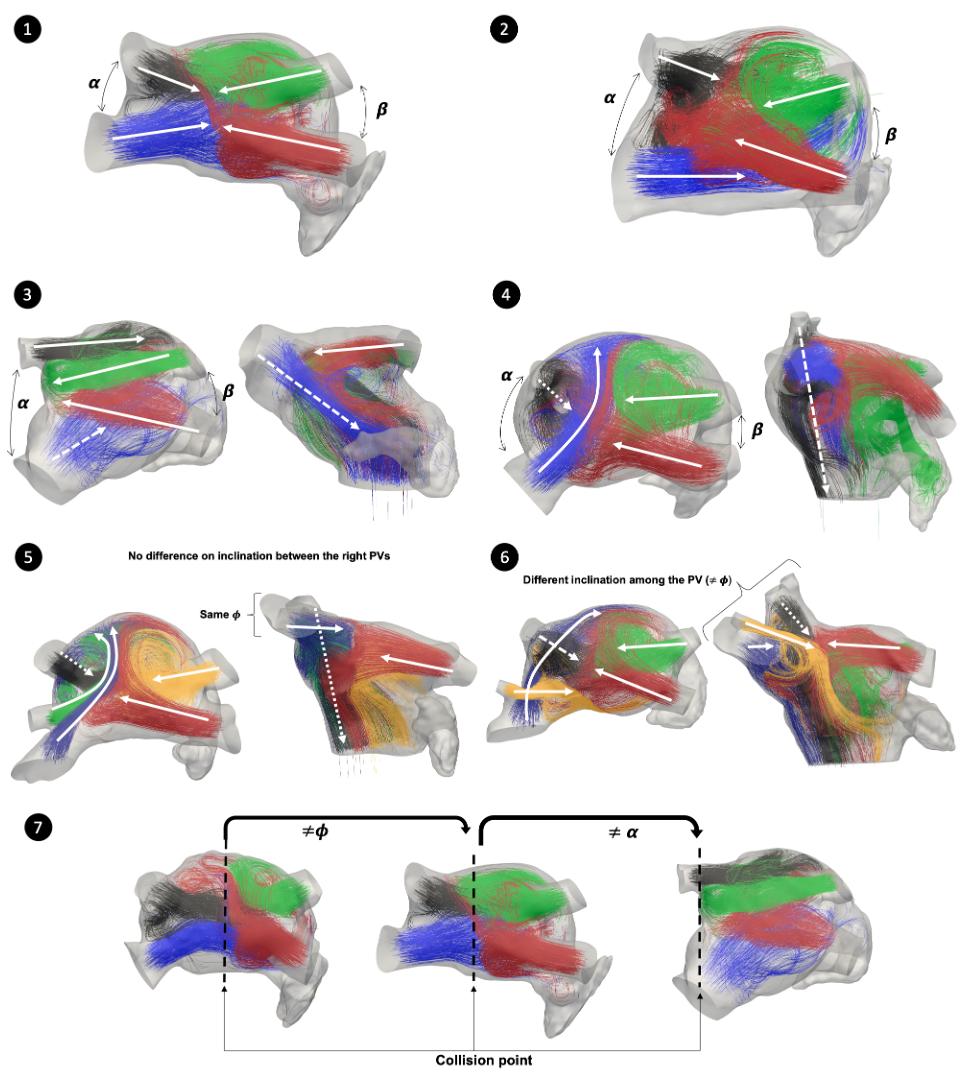}
\centering
\caption{Different scenarios of pulmonary veins (PVs) configurations and angles. 
The angle $\alpha$ is the angle between the right inferior pulmonary vein (RIPV), to and the right superior pulmonary vein (RSPV); angle $\beta$, equivalent to angle alpha but for the left superior pulmonary vein (LSPV) and left inferior pulmonary vein (LIPV); angle $\phi$ is the angle at the PVs intersection between the centre of the LA cavity and the PV centre, characterising the amount of fold between the vein and the rest of the LA..Each colour represents flow coming from the same PVs.The solid white line defines the direction of the flow coming from the PVs. The white dashed line defines the flow from the PV with the higher inclination (high $\phi$). Scenarios 1-4 had 4 PVs, while Scenario 5 and 6 had 5 and 6 PVs, respectively. In Scenario 7 the black dashed line represents where point of the flow collision.
The snapshots were taken at end diastole when the maximum velocity of the A wave is reached just before MV closing.} \label{fig2}
\end{figure}

 Figure 2 illustrates the anatomical reasons why the left side of the LA contributed with the most of the flow coming inside the LAA. When the flow coming from the left PVs collided at the superior part of the LA, it then went directly to the LAA ostium (interface with LA), which most of the times was located just under the LSPV. Actually, we realised that the angle generated between the PVs of the same side ($\alpha$, the angle between the LSPV and LIPV and $\beta$, the angle between the RSPV and RIPV in  Figure 2 in the case of 4PVs) had a considerable effect on blood flow patterns. Additionally, it was observed that the orientation of the right side PVs, their inclination with respect to the LA, varied substantially among the studied population, whereas it was quite stable for the left side PVs.
 
Scenario 1 shown in Figure 2 was the most common configuration of PVs and blood flow patterns. In  LA with 4 pulmonary veins, the collision between PVs flows usually took place at centre of the LA. If the angle between the right PVs ($\alpha$) increased, the right side flow could go laterally through the sides of the LA (Scenarios 2 and 3 in Figure 2), reaching the LAA. In our cohort, the angle between left PVs ($\beta$) did not vary as much as $\alpha$. Additionally, the inclination ($\phi$ in Figures 1 and 2) of the right PVs with respect to the LA and other PVs determined whether the blood went vertically towards the MV or towards the centre of the LA (dashed lines in Figure~\ref{fig2}). Thus, the right side PVs configuration was the main determinant factor for the whole LA haemodynamics. 

\begin{table}[t]
    \centering
    \resizebox{9cm}{!}{
    \begin{tabular}{|c||c|c|}
    \hline
  &Control &Thrombotic\\
 \hline
 3 PVs (cases/LAA volume) & 2 / 7.68 mL & 0  \\
 4 PVs (cases/LAA volume)& 11 / 12.33 mL   & 12 / 16.5 mL \\
 5 PVs (cases/LAA volume)& 9 / 10.45 mL  &  10 / 12.33 mL \\
6 PVs (cases/LAA volume)& 4 / 9.92 mL & 2 / 21.45 mL\\
 7 PVs (cases/LAA volume)& 1 / 11.27 mL & 1 / 14.95 mL\\
 \hline
 LAA volume (std) & 11.10 mL (4.20) &\textbf{15} mL (10.5)\\
 LAA/LA ratio (std)& 6.55 (2.7) & 8.11 (2.83)\\
 \hline
 \begin{tabular}{c} Flow retained  \\in LAA,all shapes (std)
 \end{tabular} & 11.57\% (9.25) & 9.68\% (11.54)  \\
\begin{tabular}{c}  Flow retained\\in LAA, CW shapes (std)\end{tabular} & 15\% (11.11)   & 2.12\% (2) \\
\begin{tabular}{c}  Flow retained  \\in LAA, non-CW shapes (std) \end{tabular} &  6.69\% (1.87)  & \textbf{8.81\%} (5.87) \\
 \hline

    \end{tabular}}
    \caption{Distribution of cases in the control and thrombotic groups based on the number of pulmonary veins (PVs), together with volumes of the left atrial appendage (LAA) and estimations of retained flow for chicken wing (CW) and non-CW LAA morphologies. Values in brackets correspond to the standard deviation. Significative differences are highlighted in bold.} 
    \label{tab:1}
\end{table}

Cases with 5 PVs shared some common morphological traits: 1) the right inferior pulmonary vein (RIPV) usually was more inclined ($\phi$ in Figure 2, and Scenario 5); and 2) the presence of the right central pulmonary vein (RCPV) shifted the RSPV be shifted towards a more transverse position. The haemodynamics in the LA in these cases were mainly influenced by the $\alpha$ and $\phi$ angles (Scenarios 5 and 6). If the RSPV were shifted to the right and pushed to a more transverse position with respect to the main LA body, the flow crossed the LA, preventing the remaining right PVs flows to collide with the left ones (Scenario 5 in Figure 2). On the other hand, if the RCPV was located in a higher position than the RSPV,RCPV was the one colliding with the left PV ones, thus RSPV flow going under RCPV one (Scenario 6). If the right PVs were very inclined, the collision point was shifted towards the right side; in the case that $\alpha$ was also large, then the collision was clearly shifted to the left side;the flow from the PVs could then reach the walls from the right side (Scenario 7 in Figure 2). Blood flow patterns observed in cases with 6 and 7 PVs were qualitatively similar to 5 PVs (Scenarios 5 and 6 in Figure 2). 

The computed values of the angles $\alpha$, $\beta$, $\phi$ and the longitudes $L_{\alpha}$ and $L_{\beta}$ are reported in Table~\ref{tab:2} for control and thrombotic groups. In general, we did not found clear relationships  between those angles and the formation of thrombi. The only remarkable difference was observed in the angles $\gamma$, $\alpha$ being larger in patients who suffered a thrombotic event and $\phi_{1}$ and $\phi_{LAA}$ being smaller smaller in patients who suffered a thrombotic event. Similarly $\phi_{LAA}$ and  $\gamma$ were also different between the control and the stroke group in patients who had 4 PVs and 6 PVs. Patients with 5 PVs who suffered stroke had a smaller $\phi_{4}$ in comparison with the control group (Table~\ref{tab:3}).

 There were not direct correlations between the risk of thrombotic event and the number of pulmonary veins (Table~\ref{tab:1}). Stroke/thrombus patients with six pulmonary veins had a lower ratio of flow inside the LAA than non-stroke cases, but differences between the LA volumes in both groups were very large and the number of samples too small (4 vs 2 cases) to draw any conclusion. The LA volume and LAA/LA ratio were higher in patients who suffered from a thrombotic event. 
Studying the LAA shape, we could find them well distributed between thrombus vs non-thrombus cases, except for the chicken wing (CW) morphologies: four chicken wings were found in the stroke group (interestingly, the four of them having more than 4 pulmonary veins, which is the most common configuration), whereas 9 where in the control group. Nevertheless, we did not found flow or anatomical differences in thrombus vs non-thrombus cases with CW morphologies. As for the relation of blood flow in the LAA with the stroke/thrombus event, fluid simulations did not show  differences between control and thrombogenic groups. Nevertheless, when analysing CW vs non-CW separately, the non-CW patients with stroke had a poorer washing than controls (see Table~\ref{tab:1}).

\begin{table}[h]
\centering
\begin{tabular}{|c||c|c|c|c|}
\hline
& \multicolumn{2}{c|}{\textbf{Systole}} & \multicolumn{2}{c|}{\textbf{Diastole}} \\

& Control & Thrombotic & Control & Thrombotic\\
\hline \hline
$\alpha$ & 50.0 (13.1) & 46.5 (11.9) & \textbf{42.5 (11.6)} & \textbf{52.5 (13)} \\
\hline
$\beta$ & 82.2 (15.7) & 80.3 (10.4)& 87.5 (16.7) & 76.6 (13.9) \\
\hline
$\alpha/\beta$ & 0.6 (0.2) & 0.6 (0.2) & \textbf{0.5 (0.2)} & \textbf{0.7 (0.2)} \\
\hline
$L_\alpha$ & 22.0 (4.8) & 21.1 (6.4) & \textbf{23.8 (4.6)} & \textbf{20.2 (2.8)} \\
\hline
$L_\beta$ & 19.5 (4.5) & 17.9 (6.0) & 19.6 (4.2) & 20.2 (3.4) \\
\hline
$L_\alpha/L_\beta$ & 1.15 (0.2) & 1.2 (0.2) & \textbf{1.2 (0.1)} & \textbf{1.0 (0.1)} \\
\hline
$\phi_1$ & 4.7 (1.8) & 5.3 (3.1) & \textbf{6.1 (2.4)} & \textbf{4.6 (2.9)} 
\\ 
\hline
$\phi_2$ & 4.8 (2.0) & 4.1 (2.8) & 3.4 (2.0) & 3.1 (1.5)
\\ 
\hline
$\phi_3$ & 15.0 (5.5) & 13.3 (5.3) & 13.3 (4.6) & 13.4 (5.7) 
\\ 
\hline
$\phi_4$ & 18.7 (5.8) & 19.1 (5.3) & 19.1 (3.7) & 17.5 (5.1)
\\ 
\hline
$\phi_{LAA}$ & 12.8 (6.4) & 13.4 (3.8) & \textbf{16.1 (5.7)} & \textbf{10.0 (3.9)}
\\
\hline
$\gamma$ & 138.0 (15.1) & 142.5 (18.4) & \textbf{141.6 (10.0)} & \textbf{152.4 (12.3)}
\\
\hline
   
\end{tabular}
 \caption{Morphological descriptors (e.g. angles) from the skeleton representation of the left atria, including its appendage (LAA), for the control and thrombogenic cases. Values are reported as the average and standard deviation (in brakets). Most significant differences are highlighted in bold.The angle $\alpha$ is the angle between the right inferior pulmonary vein (RIPV), to and the right superior pulmonary vein (RSPV); angle $\beta$, equivalent to angle alpha but for the left superior pulmonary vein (LSPV) and left inferior pulmonary vein (LIPV); $\alpha/\beta$; $L_{\alpha}/L_{\beta}$ being their respective ratios; angle $\phi$ is the angle at the PVs intersection between the centre of the LA cavity and the PV centre, characterising the amount of fold between the vein and the rest of the LA; $\phi_{LAA}$ measures the LAA bending angle described at the middle point in the LAA with respect to its ostium and its tip; $\gamma$ measures the angle between the LSPVs and the LAA.   } 
 \label{tab:2}
\end{table}

\begin{table}[t]
    \centering
    \resizebox{12cm}{!}{
\begin{tabular}{|c||c|c|c|c|c|c|}
\hline
& \multicolumn{2}{c|}{\textbf{4 PVs}} & \multicolumn{2}{c|}{\textbf{5 PVs}} & \multicolumn{2}{c|}{\textbf{6 PVs}}\\
& Control & Thrombotic
& Control & Thrombotic
& Control & Thrombotic \\
\hline \hline
$\alpha$ & 47.1(16.8) & 50.3(15.1) & 48.5(9.9) & 47.1(13.3) & 50.7(6.0) & 50.8(8.4)\\
\hline
$\beta$ & 85.1(13.3) & 78.8(11.5) & 83.2(7.7) & 78.9(13.9) & 81.2(21.7) & 83.0(5.7) \\
\hline
$\alpha/\beta$ & 0.57(0.24) & 0.63(0.16) & 0.58(0.11) & 0.63(0.30) & 0.66(0.20) & 0.61(0.05) \\
\hline
$L_\alpha$ & \textbf{24.3(4.2)} & \textbf{20.9(4.3)} & 22.5(5.5) & 21.0(3.0) & \textbf{18.6(1.8)} & \textbf{27.0(0.2)} \\
\hline
$L_\beta$ & 20.8(4.8) & 19.7(3.9) & 19.3(3.1) & 18.5(3.9) & \textbf{18.4(2.5)} & \textbf{23.4(2.9)} \\
\hline
$L_\alpha/L_\beta$ & \textbf{1.19(0.19)} & \textbf{1.06(0.16)} & 1.16(0.16) & 1.17(0.27) & 1.03(0.21) & 1.16(0.15) \\
\hline
$\phi_1$ & 4.9(2.4) & 4.4(2.7) & 6.3(1.5) & 5.9(2.7) & \textbf{4.1(2.7)} & \textbf{1.5(0.6)} \\ 
\hline
$\phi_2$ & 3.9(2.4) & 3.5(1.2) & 4.3(2.2) & 3.4(2.0) & \textbf{5.2(1.5)} & \textbf{2.3(0.2)} \\ 
\hline
$\phi_3$ & 14.0(3.7) & 12.1(5.7) & 14.4(5.0) & 13.5(4.7) & 16.4(8.9) & 15.2(4.4) \\ 
\hline
$\phi_4$ & 18.3(4.8) & 18.0(5.3) & \textbf{21.4(5.7)} & \textbf{16.7(3.0)} & 17.9(3.8) & 22.1(1.6)\\ 
\hline
$\phi_{LAA}$ & \textbf{16.7(5.8)} & \textbf{11.9(3.9)} & 10.1(1.8) & 10.6(2.7) & \textbf{12.2(4.9)} & \textbf{9.9(0.6)} \\
\hline
$\gamma$ & \textbf{129.8(17.8)} & \textbf{143.4(15.2)} & 123.7(25.4) & 120.9(30.5) & \textbf{133.5(12.2)} & \textbf{139.3(8.8)} \\
\hline
\end{tabular}}
 \caption{Morphological descriptors for control and thrombotic groups, separated by the number of pulmonary veins (PVs). Values are reported as average and standard deviation (in brackets). Significant values are highlighted in bold. Morphological descriptors for control and thrombotic groups, separated by the number of pulmonary veins (PVs). Values are reported as average and standard deviation (in brackets). Significant values are highlighted in bold. $\alpha$ is the angle between the right inferior pulmonary vein (RIPV), to and the right superior pulmonary vein (RSPV); angle $\beta$, equivalent to angle alpha but for the left superior pulmonary vein (LSPV) and left inferior pulmonary vein (LIPV); $\alpha/\beta$; $L_{\alpha}/L_{\beta}$ being their respective ratios; angle $\phi$ is the angle at the PVs intersection between the centre of the LA cavity and the PV centre, characterising the amount of fold between the vein and the rest of the LA; $\phi_{LAA}$ measures the LAA bending angle described at the middle point in the LAA with respect to its ostium and its tip; $\gamma$ measures the angle between the LSPVs and the LAA.} 
 \label{tab:3}
\end{table}

%\begin{figure}
%\includegraphics[width=\textwidth]{figure3.png}
%\caption{Studied cases more likely to flow stagnation are presented. Top panel: cases have an alignment between the left superior pulmonary vein and the LAA (red line). Bottom panel: cases have a great angulation of the LAA, with its tips pointing toward the mitral valve. } \label{fig3}
%\end{figure}

\section{Discussion and conclusions}
The presented study analyses how the PVs configuration influences LA haemodynamics, being the first time these factors are related to thrombus risk due to potential LAA flow stagnation. In order to do so, we have created the largest database of LA fluid simulations in the literature so far; until now, less than 10 different real LA anatomies were processed in each study, a number substantially lower than the 52 cases employed in this work. In general, LA haemodynamics were mostly influenced by morphological characteristics  of the right PVs. Their angles and inclinations determined the point of collision of the multiple PV flows in the LA. These were key to understand how blood patterns evolved evolved and traversed the whole LA main cavity, in particular how they were reaching the LAA (key for thrombus formation risk estimation). 
According to the obtained results, the number of the PVs is not directly related to the probability to suffer a thrombotic event. In agreement with the literature, significant differences were found between LAA volume and LAA/LA ratio between control and thrombotic cases. Also, CW morphologies were less likely to generate thrombus. Nevertheless, in the few CW cases with stroke, the number of PVs was larger than four, which is the most frequent configuration. Despite not finding a strong relationship between thrombus formation and the number of PVs or LAA flow washing in the whole population, the independent analysis of the non-chicken wing cases showed a better LAA flow washing in the control vs thrombotic groups. Finally, some PV angle values were significally different in controls vs thrombotic cases ($\gamma$, $\alpha$,$\phi_{1}$, $\phi_{LAA}$ and $\phi_{4}$) although the relation was not strong. These results complement the qualitative results obtained with the streamlines where RSPVs ($\phi_{1}$ and $\alpha$) had a great influence on LA haemodynamics. The angle $\phi_{LAA}$ was also mentioned by \cite{Yaghi2020} as possible new risk factor of thrombus formation in LAA.

One limitation in our study was the absence of any Doppler data acquired during an AF event, therefore, the A wave was present in the boundary conditions of the fluid simulations. However, an AF scenario was replicated by imposing a pressure curve at the mitral valve as well as omitting the LA radial (i.e., active) contraction. In the future, we will explore the use of pathlines for a better assessment of LA flow patterns.

\subsubsection{Funding} 
This work was supported by the Agency for Management of University and Research Grants of the Generalitat de Catalunya under the the Grants for the Contracting of New Research Staff Programme - FI (2020 FI\_B 00608) and the Spanish Ministry of Economy and Competitiveness under the Programme for the Formation of Doctors (PRE2018-084062), the Maria de Maeztu Units of Excellence Programme (MDM-2015-0502) and the Retos Investigación project (RTI2018-101193-B-I00). Additionally, this work was supported by the H2020 EU SimCardioTest project (Digital transformation in Health and Care SC1-DTH-06-2020; grant agreement No. 101016496) and the European project PARIS (ID35).

\bibliographystyle{splncs03_unsrt}
\bibliography{References.bib}

\end{document}